\documentclass[12pt]{article}
\usepackage{amssymb,amsmath}

\def \Cor {\mathfrak{C}}

\def \DD {\mathcal{D}}
\def \FF {\mathcal{F}}
\def \HH {\mathcal{H}}

\def \CCC {\mathbb{C}}
\def \NNN {\mathbb{N}}

\def \tr {\mathrm{Tr}}
\def \ker {\mathrm{Ker}}

\def \bs {\mathbf{s}}

\def \equals {\ = \ }
\def \plus {\ + \ }

\def \quasifree {\Gamma}

\newtheorem{proposition}{Proposition}
\newtheorem{corollary}{Corollary}

\title{Properties of nonfreeness: an entropy measure \\ of electron correlation}

\author{
Alex D.\  GOTTLIEB \footnote{ Wolfgang Pauli Institute,
Nordbergstr. 15, A--1090 Wien, Austria (alex@alexgottlieb.com).
} \\
and Norbert J.\ MAUSER \footnote{ WPI c/o Fak. f.\ Math., Univ.
Wien, Nordbergstr. 15, A--1090 Wien, (mauser@courant.nyu.edu). } }

\date{}

\begin{document}
\maketitle

\begin{abstract}
``Nonfreeness" is the (negative of the) difference between the von Neumann entropies of 
a given many-fermion state and the free state that has the 
same $1$-particle statistics.  
It also equals the {\it relative} entropy of the two states in question, 
i.e., it is the entropy of the given state relative to the corresponding 
free state.  
The nonfreeness of a {\it pure} state is the same as its ``particle-hole symmetric correlation entropy", 
a variant of an established measure of electron correlation.  
But nonfreeness  is also defined for {\it mixed} states, and this allows 
one to compare the nonfreeness of subsystems to the nonfreeness of the whole.  
Nonfreeness of a part does not exceed that in the whole; nonfreeness is 
additive over independent subsystems; and nonfreeness is superadditive 
over subsystems that are independent on the $1$-particle level.
\end{abstract}

The word ``correlation" in the context of many-electron systems is somewhat 
overcharged and ambiguous, except when used in the expression ``correlation energy," 
where it refers to the difference between the energy of the true ground state of 
a many-electron system and the energy of the Hartree-Fock approximation.   
Usually, ``correlation effects" refers to properties of a many-electron state that 
cannot be explained if its wavefunction is the single Slater determinant obtained 
by the Hartree-Fock method; 
but sometimes, even those classical 
statistical correlations  rendered necessary by the very antisymmetry of fermion 
wavefunctions, as manifested, 
for example, in the phenomenon of the ``Fermi hole," are also described as 
``Fermi correlations" \cite{Kutzelnigg}.    
Here we understand ``correlation" in the former sense, and interest ourselves in 
measures of correlation that quantify the degree to which a given many-electron 
state can be distinguished from states pertaining to ``free" (i.e., noninteracting) 
particles, e.g., states whose wavefunctions are Slater determinants.    
Such measures of electron correlation depend only on the given state, 
without reference to any extrinsic Hamiltonian, and therefore without reference 
to any prescribed ``correlation energy."  
Several of these correlation measures are functions of the eigenvalues of the 
$1$-particle density matrix (1PDM).  This class includes the ``nonidempotency" 
of the 1PDM \cite{LichtnerGriffin,Ziesche}, the ``degree of correlation" 
\cite{GrobeRzazewskiEberly}, and the ``correlation entropy" 
\cite{Ziesche,EsquivelEtAl,GersdorfEtAl,ZiescheEtAl,ZiescheEtAl2}.  
Other correlation measures use the $1$-particle and $2$-particle position 
and momentum distributions, and quantify correlation in terms of the usual 
statistical correlation \cite{Kutzelnigg} or in terms of information 
\cite{GadreEtAl,GuevaraSagarEsquivel,SagarGuevara}. 

In this article we prove that a version of the correlation entropy mentioned above, namely, the ``particle-hole symmetric correlation entropy" of Ref.~\cite{Gori-GiorgiZiesche}, is naturally extended to the domain of mixed states, so that the resulting measure of electron correlation, which we call 
``nonfreeness", 
behaves well when one considers of subsystems of a given many-electron system.  The state of a subsystem of electrons --- e.g., the electrons of a 
$\mathrm{CH}_4$ molecule that may be found within $0.5$\AA\  of the 
carbon nucleus --- is generally a mixed state of variable particle number.  
Using nonfreeness allows one to speak of the ``correlation" in the subsystem, or rather in the subsystem's {\it state}, and compare it to the correlation in the electronic state of the whole molecule.  It will be seen that the nonfreeness of the state of the subsystem cannot exceed the nonfreeness of the state of the whole system.  We will also show that nonfreeness is superadditive over disjoint substems, if the states of the systems are independent on the $1$-particle level.  These properties of nonfreeness, we feel, make it a superior measure of electron correlation.

Nonfreeness is defined here in the spirit of our published Letter \cite{Us}, where we proposed to quantify electron correlation  
by comparing a given many-electron state to the free state 
that shares the same $1$-particle statistics.  To compare the states, we use relative entropy:        
the {\bf nonfreeness of a general many-electron state is the entropy of that 
state relative to the free state with the same $1$-particle correlation 
operator}.   The nonfreeness of a state $\omega$ will be denoted by $\Cor(\omega)$.

This article concentrates on {\it finite} electronic systems, represented by antisymmetric $n$-electron wavefunctions, or (more generally) by 
density operators $\Delta$ on the fermion Fock space $\FF_{\HH}$ with finite expected particle number.  
Proposition~1 below gives a useful formula for the nonfreeness of a state of this kind.  Propositions~2 and 3 are stated in the same context of finite systems, but they hold true for infinite systems as well.  Nonfreeness for infinite systems is discussed in the concluding section.  

Let $\DD$ denote the set of all density operators on  $\FF_{\HH}$  such that 
\begin{equation}
\label{D}
\Delta N = N\Delta \qquad \hbox{and} \qquad \tr(\Delta N) < \infty\ ,
\end{equation}
where $N$ is the number operator defined in formula (\ref{number}) below.    
For $\Delta \in \DD$, let $\gamma_{\Delta}$ denote the $1$-particle statistical 
operator $\gamma_{\Delta}$ defined in formula (\ref{1PDM}) below, 
and let $\quasifree_{\gamma_{\Delta}}$ denote the density operator of the unique  free state 
with $1$-particle statistical operator $\gamma_{\Delta}$.   
Proposition~\ref{p1} states that the nonfreeness of $\Delta$ is the difference between the von Neumann entropy 
of $\quasifree_{\gamma_{\Delta}}$ and the von Neumann entropy 
of $\Delta$, provided that the former entropy is finite.    
In such cases $\Cor(\Delta)$ equals a simple functional  
of the natural occupation probabilities, minus the von Neumann entropy 
of $\Delta$.  

Propositions~\ref{monotonicity} and \ref{superadditivity} concern the way the 
nonfreeness of a many-electron system relates to the nonfreeness of its subsystems.  
Suppose that $\HH_1$ is a closed subspace of the $1$-particle Hilbert space $\HH$, so that $\HH = \HH_1 \oplus \HH_2$.  The Hilbert space of the whole system is $\FF_{\HH}$, and the fermion Fock space over $\HH_1$ and $\HH_2$ are the Hilbert spaces of the subsystems under consideration.  A state $\omega$ of the whole system induces states $\omega_1$ and $\omega_2$ of the subsystems.   Proposition~\ref{monotonicity} states 
that $\Cor(\omega) \le \Cor(\omega_1)$.  Proposition~\ref{superadditivity} states that $\Cor(\omega_1) + \Cor(\omega_2) \le \Cor(\omega)$ if the subsystems are ``independent on the $1$-particle level."

\newpage

\noindent {\bf The nonfreeness of wavefunctions and density operators:}

The nonfreeness of a pure $n$-electron state is an entropy-type function of the natural occupation probabilities.  
Given a normalized antisymmetric wavefunction  $\psi(x_1,\ldots,x_n)$ representing an $n$-electron state, let $\gamma_{\psi}$ denote the $1$-particle ``reduced density matrix," normalized to have trace $n$: this is an operator with integral kernel
\[
     \gamma_{\psi}(x,y) \equals n \int dz_n  \cdots \int  dz_2  \ \psi(x,z_2,\ldots,z_n) \overline{\psi(y,z_2,\ldots,z_n)} \ .
\]
The eigenvalues of the operator $\gamma_{\psi}$ lie between $0$ and $1$ and are known as ``natural occupation probabilities."  We define the ``nonfreeness of $\psi$" by 
\begin{equation}
\label{CorPure}
    \Cor(\psi)  \equals   - \sum p_j \log p_j - \sum (1-p_j) \log(1-p_j) 
\end{equation}
where the $p_j$ are the natural occupation probabilities for $\psi$.  This is the entropy of the  free state built from the spectral decomposition of $\gamma$ (viz. Section ~9.4.1 of \cite{AlickiFannes}).  $\Cor(\psi)=0$ if and only if $\psi$ is a Slater determinant.  
The first sum in (\ref{CorPure}) is known as the ``correlation entropy" \cite{Ziesche,GersdorfEtAl,ZiescheEtAl}.   Formula (\ref{CorPure}) itself has been introduced and applied in Ref.~\cite{Gori-GiorgiZiesche}, where it is called ``particle-hole symmetric correlation entropy."

The nonfreeness functional $\Cor$ extends to mixed states of variable particle number so that  free states, and only such states, have $0$ nonfreeness.  To discuss such states, we need to recall the concept of fermion Fock space and the second quantization of operators.  Let $\HH$ denote the $1$-particle Hilbert space, and let 
\[
     \FF_{\HH} \equals \CCC \oplus \HH \oplus (\HH \wedge \HH) \oplus (\HH \wedge \HH \wedge \HH) \oplus \cdots
\]
denote the fermion Fock space over $\HH$.  The first summand on the right hand side $(\CCC)$ is spanned by the vacuum vector 
\[
    \Omega \ \equiv\  {\bf 1}_{\CCC} \oplus 0 \oplus 0 \oplus \cdots.
\]
For any $h \in \HH$, the corresponding creator $a^{\dagger}_h$ 
and annihilator  $a_h$ are represented by bounded operators on $\FF_{\HH}$.   
The creator is the adjoint of the corresponding annihilator; 
any two annihilators $a_g$ and $a_h$ anticommute; and 
\begin{equation} \label{anticomm}
   a^{\dagger}_g a_h + a_h a^{\dagger}_g \equals 
\langle h | g \rangle {\mathrm I}_{\FF_{\HH}}.
\end{equation}
Let $f_1,f_2,\ldots $ be any ordered orthonormal basis of $\HH$, and set $a_j \equiv  a_{f_j}$.   
The number operator on $\FF_{\HH} $ is 
\begin{equation}
\label{number}
 N \equals \sum a^{\dagger}_j a_j 
\end{equation}
(the operator so defined does not really depend on the choice of ordered orthonormal basis).  
For any finite subset $\bs$ of $\NNN = \{1,2,3,\ldots\}$, set 
\[
  a_{\bs} \equals a_{s_1}a_{s_2} \cdots a_{s_n} 
  \qquad \hbox{and} \qquad 
    a^{\dagger}_{\bs}  \equals a^{\dagger}_{s_n}\cdots a^{\dagger}_{s_2} a^{\dagger}_{s_1} \ ,
\]  
where $s_1 < s_2 < \ldots < s_n$ are the elements of $\bs =\{s_1, s_2, \ldots, s_n\}$, indexed in increasing order.  
When $\bs$ is the empty subset of $\NNN$, let $a_{\bs}$ and $a^{\dagger}_{\bs}$ denote the identity operator on $\FF_{\HH} $.  
The set of all vectors $a^{\dagger}_{\bs}\Omega $ in $\FF_{\HH} $, where $\bs$ ranges over finite subsets of $\NNN$, including the empty set, is an orthonormal basis of the fermion Fock space.   For any $\bs \subset \NNN$, the operator $a^{\dagger}_{\bs}a_{\bs}$ is an orthogonal projector.

For any density operator $\Delta$ on the fermion Fock space, 
the $1$-particle statistical operator $\gamma_{\Delta}$ is defined such that 
\begin{equation}
\label{1PDM}
     \langle h | \gamma_{\Delta} g \rangle \equals  \tr(\Delta a^{\dagger}_g a_h ) 
\end{equation}
for any $g,h \in \HH$, where $a^{\dagger}_x$ denotes the creator of $x$ on $\FF_{\HH}$.   
$\gamma_{\Delta}$  is a Hermitian operator whose spectrum is contained in the interval $[0,1]$.  
The average particle number is $\tr(\Delta N)=\tr(\gamma_{\Delta})$, 
so $\gamma_{\Delta}$ has finite trace if and only if the average particle number is finite.  

Recall the class $\DD$ of all density operators on  $\FF_{\HH}$ that satisfy conditions (\ref{D}) above.
For any  Hermitian operator $\gamma$ on $\HH$ with finite trace and all eigenvalues in the interval $[0,1]$, there exists a unique free state, represented by a density operator $\quasifree_{\gamma} \in \DD$, such that $\gamma_{\quasifree_{\gamma}}=\gamma$.   
The spectral decomposition of $\quasifree_{\gamma}$ can be constructed explicitly as follows \cite{AlickiFannes}.    
Given $\gamma$, let $p_1 \ge p_2 \ge \cdots$ be the list of positive eigenvalues of $\gamma$, 
and let $f_1,f_2,\ldots$ be a list of corresponding eigenvectors.   
Set $a_j \equiv  a_{f_j}$, and define the operators $a^{\dagger}_{\bs}$ as above for finite subsets $\bs =\{s_1,\ldots,s_n\} \subset \NNN$.  Then
\begin{equation}
\label{spectralQuasifree}
    \quasifree_{\gamma} \equals \sum_{\bs}    p(\bs) \ |  a^{\dagger}_{\bs}\Omega \rangle \langle a^{\dagger}_{\bs}\Omega |
\end{equation}
with 
\begin{equation}
\label{weightQuasifree}
    p(\bs) \equals \ \prod_{i \in \bs} p_i \prod_{j \notin \bs} (1-p_j) \ .
\end{equation}
$\quasifree_{\gamma_{\Delta}}$ is the (density operator of the) free state with the same $1$-particle statistics as $\Delta$.   

Let $ S(X) $ denote the von Neumann entropy $ -\tr(X\log X) $ of a density operator $X$, and let 
$ S(X|Y) $ denote the relative entropy $ -\tr(X(\log X - \log Y))$ of two density operators $X$ and $Y$  
 \cite{OhyaPetz,AlickiFannes}.  The von Neumann entropy and relative entropy of density operators are always nonnegative, but may equal $+\infty$.  
In particular, for density operators $X$ and $Y$,  $ S(X|Y) $ 
is defined to equal $+\infty$ if the kernel of $X$ is not contained in the kernel of $Y$. 
For $\Delta \in \DD$, we define the ``nonfreeness of $\Delta$" as
\begin{equation}
\label{Cor}
\Cor(\Delta) \equals     S(\Delta | \Gamma_{\gamma_{\Delta}} ) \ .
\end{equation}      
This equals $0$ if and only if $\Delta =  \quasifree_{\gamma_{\Delta}}$, and it is never negative (though it may equal $+\infty$).  
In case the state is a pure state given by a $n$-electron wavefunction $\psi$, the corresponding density operator is 
\[
    \Delta \equals
    0 \oplus \cdots \oplus 0
    \stackrel{n-particles}{\oplus\ |\psi \rangle \langle \psi |\ \oplus }  0
    \oplus 0 \oplus \cdots\ 
\]
and it may be verified that $ \Cor(\Delta) = \Cor(\psi)$.

\begin{proposition}
\label{p1}
Suppose $\Delta \in \DD$ and 
\begin{equation}
\label{piaoliang}
   - \sum p_j \log p_j - \sum (1-p_j) \log(1-p_j) \ < \ \infty\ ,        
\end{equation}
where $p_1,p_2,\ldots$ are the eigenvalues of $\gamma_{\Delta }$.  Then 
\begin{equation}
\label{meili}
   \Cor(\Delta) \ \equiv\ S( \Delta | \quasifree_{\gamma_{\Delta}} ) \equals - \sum p_j \log p_j - \sum (1-p_j) \log(1-p_j)  - S(\Delta)\ .
\end{equation}
\end{proposition}

\noindent The proof of Proposition \ref{p1} appears in Appendix A. 

Proposition \ref{p1} means that the nonfreeness of a state represented by a 
density operator  $\Delta \in \DD$ is the amount that the von Neumann entropy 
of $\Delta$ falls short of the entropy of the corresponding 
 free state $\quasifree_{\gamma_{\Delta}}$, which is the largest 
von Neumann entropy possible for all states that have the same
$1$-particle operator $\gamma_{\Delta}$. 
\begin{corollary}
\label{c0}
Under the hypotheses of Proposition \ref{p1}, the von Neumann entropy of a state satisfies
\[
     S(\Delta) \ \le \  - \sum p_j \log p_j - \sum (1-p_j) \log(1-p_j) \ ,
\]
with equality if and only $\Delta =  \quasifree_{\gamma_{\Delta}}$. 
\end{corollary}
\noindent{Proof:} \qquad  
This follows from equation (\ref{meili}) and the observation that $S( \Delta | \quasifree_{\gamma_{\Delta}} )$ equals $0$ if 
$\Delta = \quasifree_{\gamma_{\Delta}} $, otherwise $S( \Delta | \quasifree_{\gamma_{\Delta}} )$ is strictly positive.
\hfill $\square$

\begin{corollary}
\label{c1}
Suppose $\Delta \in \DD$ and $\mathrm{rank}(\gamma_{\Delta} ) =k$.  Then  $\Cor(\Delta) \le k $.
\end{corollary}

\noindent{Proof:} \qquad  
Since $\gamma_{\Delta}$ has rank $k$, the von Neumann entropy of $S(\quasifree_{\gamma_{\Delta}}) \le k < \infty$, and Proposition~\ref{p1} implies that $\Cor(\Delta) \le S(\quasifree_{\gamma_{\Delta}}) \le k$.     
\hfill $\square$

The upper bound in Corollary~\ref{c1} is not always attained.  Indeed,  $\Cor(\Delta) = 0 $ if  $\mathrm{rank}(\gamma_{\Delta} ) = 1$.  Also, $\Cor(\Delta) \le 1 $ if $\mathrm{rank}(\gamma_{\Delta} ) = 2$ (this can be shown using formula (\ref{rank 2}) of Appendix C).  However,  if $2m > 2$ is an even number, then there do exist states $\Delta$ such that $\mathrm{rank}(\gamma_{\Delta} ) = 2m$ and $\Cor(\Delta) = 2m$.  For example, let 
$\{ \phi_1,\ldots,\phi_m, \psi_1, \ldots, \psi_m\}$ be an orthonormal set in $\HH$, and let $\Phi$ and $\Psi$ denote Slater determinants in $\phi_1, \ldots, \phi_m$ and $\psi_1, \ldots, \psi_m$, respectively.  Then 
\[
     \Cor\big( \tfrac{1}{\sqrt{2}} \Phi + \tfrac{1}{\sqrt{2}} \Psi \big) \equals 2m \ ,
\]
that is, the nonfreeness of $ \tfrac{1}{\sqrt{2}} ( \Phi +  \Psi)$ attains the maximum possible for states of rank $2m$.

\newpage

\noindent {\bf Nonfreeness and subsystems:}

The electronic state of a molecule determines the properties of any ``subsystem" of the molecule's electrons.  
In this section we will consider subsystems of a special form, including subsystems consisting of precisely those electrons 
that occupy a given region of space or a given bond orbital.  
If the electronic state of the molecule is given by a Slater determinant, then any subsystem is in a  free state.    But if the molecule is in a ``correlated" state, the subsystem will typically be in a correlated state too.   One would expect there to be less ``correlation" in the subsystem than there is in the whole molecule, and indeed, nonfreeness behaves this way: it is monotone with respect to consideration of subsystems.  The ``monotonicity" of the nonfreeness $\Cor$ is a consequence of the monotonicity of quantum relative entropy, a very important and rather deep property of quantum entropy \cite{Petz}.   The monotonicity of quantum relative entropy was first established for density operators by Lindblad \cite{Lindblad74,Lindblad75} and later for general states on von Neumann algebras by Uhlmann  \cite{Uhlmann77,OhyaPetz}.   

The kind of subsystem we consider here has the following general form.  Let $\HH_1$ be a (closed) subspace of the $1$-particle space $\HH$, and let $\HH_2$ be the complementary subspace, so that $\HH=\HH_1 \oplus \HH_2$.  
For example, if we want to consider the electronic subsystem associated to a region  $R$ of space, 
then $\HH_1$ will be the space of spin-orbitals $\psi$ such that $\psi(r,\sigma)=0$ unless $r$ lies in the region $R$. 
If we want to consider the electronic state restriced to a bond orbital $\phi$, then $\HH_1$ will be the span of $|\phi\uparrow\rangle$ and $|\phi\downarrow\rangle$.  Let $\FF_1$ and $\FF_2$ denote the Fock spaces $\FF_{\HH_1}$ and $\FF_{\HH_2}$, respectively.   
The Fock spaces $\FF_1$ and $\FF_2$ may be regarded as subspaces of $\FF_{\HH}$: 
for example, $\FF_1$ is isomorphic to the subspace of $\FF_{\HH}$ that is spanned by Slater determinants in spin-orbitals taken from $\HH_1$.  
The whole Fock space $\FF_{\HH}$ is isomorphic to $\FF_1 \otimes \FF_2 $ as follows \cite{AlickiFannes}.  Let $\{h_j\}$ be an orthonormal basis of $\HH$ such that each $h_j$ is either in $\HH_1$ or in $\HH_2$, 
 and set $S=\{ j \in \NNN: h_j \in \HH_1 \}$ and $S'=\NNN \setminus S$.  The map
\begin{equation}
\label{isomorphism}
       a^{\dagger}_{\bs \cup \bs'}\Omega  \longleftrightarrow  a^{\dagger}_{\bs}\Omega \otimes a^{\dagger}_{\bs'} \Omega 
\end{equation}
defined for $\bs \in S, \bs' \in S'$ extends to an isomorphism.  If $\Delta$ is a density operator on $\FF$, let 
\[
       \Delta_1 \equals \tr_{\FF_2}(\Delta)
\]
denote the partial trace of $\Delta$ over $\FF_2$.  This is a density operator on $\FF_1$ that we will call the ``restriction of $\Delta$ over $\HH_1$."

From definition (\ref{1PDM}) of the $1$-particle statistical operator, it follows that $\gamma_{\Delta_1}$ is the compression to $\HH_1$ of the $1$-particle statistical operator for $\Delta$, i.e.,
\begin{equation}
\label{compression1}
     \gamma_{\Delta_1}  \equals P_1 \circ  \gamma_{\Delta} \ \big|_{\HH_1}\ , 
\end{equation}
where $P_1$ denotes the orthogonal projector on $\HH$ with range $\HH_1$.

We return to the subject of nonfreeness.   For the rest of this section, all density operators are implicitly assumed to lie in the class $\DD$ (they commute with the number operator and have finite expected particle number), for we have only defined nonfreeness for density operators in this class.  However, as we discuss in the concluding remarks, the definition of nonfreeness can be generalized to apply to {\it all} many-fermion states and the results of this section remain true in the greatest generality.

\begin{proposition}
\label{monotonicity}
Let $\Delta$ be a density operator on the fermion Fock space over $\HH$ and let $\HH_1$ be a subspace of $\HH$.  
Then $\Cor(\Delta_1) \le \Cor(\Delta)$.  
\end{proposition}

\noindent{Proof:} \qquad   
If $\gamma$ is a bounded $\quasifree_{\gamma}$ is the density operator of a free state,  then 
its restriction over $\HH_1$ is also a  free state.  In fact
\begin{equation}
\label{compression2}
    (\quasifree_{\gamma})_1 \equals \quasifree_{P_1 \gamma |_{\HH_1}}\ .
\end{equation}  
By (\ref{compression2}) and (\ref{compression1}),  $(\quasifree_{\gamma_{\Delta}})_1= \quasifree_{ \gamma_{\Delta_1} }$.  Thus, the inequality 
\[
  \Cor(\Delta) \equals S(\Delta| \quasifree_{\gamma_{\Delta}}) \ \ge \  S\big( \tr_{\FF_2}(\Delta) | \tr_{\FF_2}( \quasifree_{\gamma_{\Delta}}) \big) \equals 
S\big(\Delta_1| (\quasifree_{\gamma_{\Delta}})_1 \big) \equals S\big(\Delta_1| \quasifree_{ \gamma_{\Delta_1} } \big) \equals \Cor(\Delta_1) 
\]
follows from the monotonicity of relative entropy (viz. Lemma~2 of \cite{Lindblad75}).   
\hfill $\square$

Let $\Delta$ be a density operator on the Fock space over a Hilbert space $\HH = \HH_1 \oplus \HH_2$ and let $\Delta_1$ and $\Delta_2$ denote the restrictions of $\Delta$ over $\HH_1$ and $\HH_2$.  We say that the two subsystems corresponding to $\HH_1$ and $\HH_2$ are ``independent" if $\Delta \ \widetilde{=} \ \Delta_1 \otimes \Delta_2$.  
We say they are ``independent on the $1$-particle level" if $\Delta$ and $\Delta_1 \otimes \Delta_2$ have the same $1$-particle statistical operator, or (equivalently) if $\gamma_{\Delta}=\gamma_{\Delta_1}\oplus \gamma_{\Delta_2}$ with respect to the decomposition $\HH_1 \oplus \HH_2$ of $\HH$.   The nonfreeness of independent subsystems is additive, i.e., 
\[
     \Cor(\Delta_1 \otimes \Delta_2) = \Cor(\Delta_1) + \Cor(\Delta_2)\ .
\]
  The next proposition states that the nonfreeness is {\it superadditive} if the 
subsystems are independent on the $1$-particle level.  
This follows from the superadditivity of entropy and the fact that, 
for  free states, independence on the $1$-particle level implies 
independence, i.e., $\quasifree_{\gamma_1 \oplus \gamma_2} = 
\quasifree_{\gamma_1} \otimes \quasifree_{\gamma_2}$. 
\begin{proposition} 
\label{superadditivity} 
Let $\Delta$ be a density operator on the Fock space over a Hilbert space $\HH = \HH_1 \oplus \HH_2$ and let $\Delta_1$ and $\Delta_2$ denote the restrictions of $\Delta$ over $\HH_1$ and $\HH_2$.  
If $\gamma_{\Delta} = \gamma_{\Delta_1}\oplus \gamma_{\Delta_2}$ then 
\begin{equation}
\label{super}
  \Cor(\Delta) \ \ge\ \Cor(\Delta_1) + \Cor(\Delta_2) \ .
\end{equation}
If $\Cor(\Delta) < \infty$, then equality holds in (\ref{super}) if and only if $\Delta \ \widetilde{=} \ \Delta_1 \otimes \Delta_2$.
\end{proposition}

\noindent{Proof:} \qquad   
If $\gamma_{\Delta} = \gamma_{\Delta_1} \oplus \gamma_{\Delta_2}$ then $\quasifree_{\gamma_{\Delta}} 
\ \widetilde{=}\  \quasifree_{\gamma_{\Delta_1}} \otimes \quasifree_{\gamma_{\Delta_2}}$.  
Superadditivity of relative entropy implies the inequality 
\[
   \Cor(\Delta) \equals 
   S(\Delta| \quasifree_{\gamma_{\Delta}}) 
   \equals
   S(\Delta \ | \ \quasifree_{\gamma_{\Delta_1}} \otimes \quasifree_{\gamma_{\Delta_2}}) 
   \\
   \ \ge \ S(\Delta_1 | \quasifree_{\gamma_{\Delta_1}} ) + S(\Delta_1 | \quasifree_{\gamma_{\Delta_2}} )
   \equals \Cor(\Delta_1) + \Cor(\Delta_2)\ ,
\]
with the stated conditions for equality (see Corollary~5.21 of \cite{OhyaPetz}).
\hfill $\square$

For example, consider a many-electron state in which there is a precise number of electrons in some region $R_1$ of space.  The two subsystems consisting of (i) the electrons in $R_1$, and (ii) the electrons in the complementary region $R_2$, are independent on the $1$-particle level.  Therefore the nonfreeness of the state is greater than (or equal to) the nonfreeness of the restriction of the state over $R_1$ plus the nonfreeness of the restriction of the state over $R_2$.  

More generally, Proposition~\ref{superadditivity} has the following consequence, 
whose proof appears in Appendix B.

\begin{corollary} 
Let $\FF_{\HH}$ denote the Fock space over a Hilbert space $\HH = \HH_1 \oplus \HH_2$ 
and let $N_1$ denote the number operator for the subspace $\HH_1$.  
If $\Delta$ is a density operator on $\FF_{\HH}$ 
such that $N_1\Delta = \Delta N_1 = n \Delta$ for some integer $n$, 
then
\[
    \Cor(\Delta) \ \ge \ \Cor(\Delta_1) + \Cor(\Delta_2) 
\]
where $\Delta_1$ and $\Delta_2$ denote the restrictions of $\Delta$ over $\HH_1$ and $\HH_2$.
\end{corollary}

\newpage

\noindent {\bf An example: the nonfreeness of a highly correlated state of lattice fermions}

Consider a system of $2L$ electrons, half of them of spin up and half of spin down, on a lattice of $2L$ sites, 
subject to the Hubbard Hamiltonian with an on-site repulsion of strength $U$ and with nearest neighbor hopping of strength $t$. 
 In the ``atomic" limit, where $t$ tends to $0$ while $U$ remains constant, the on-site repulsion forces there to be one electron at each lattice site.  
 Let $\HH_0$ denote the subspace of states for which there is only one electron per site.  
 The compression of the Hubbard Hamiltonian to $\HH_0$ is the zero operator to first order in $t$, but to second order in $t$ it is the Hamiltionian of the antiferromagnetic Heisenberg model.  This second order effect, called ``superexchange" by P. W. Anderson \cite{Anderson}, can be justified rigorously \cite{KleinSeitz} using degenerate perturbation theory \cite{Kato}.

Thus, in the atomic limit, the ground state of the Hubbard model tends toward the ground state of the antiferromagnetic Heisenberg model, 
{\it considered as a state of the system of lattice fermions}.  
In this state there is one electron per site, and therefore the $1$-electron density matrix is diagonal; 
moreover, every eigenvalue of the $1$-electron density matrix is equal to $1/2$ thanks to spin symmetry.
By Proposition~\ref{p1}, the nonfreeness of this state equals $4L$, 
which is the maximum possible for any many-electron state on a lattice of $2L$ sites, by Corollary~\ref{c1}.    

The nonfreeness of the restriction of the ground state over one site is also as large as possible: it equals $1$.
If there were no correlation between sites, then the total nonfreeness would be $2L \times 1 = 2L$ by Proposition~\ref{superadditivity},
but in fact the nonfreeness is $4L$, not $2L$.  This means that half of the nonfreeness of the state is due to correlations between different sites.

\newpage

\noindent {\bf Concluding remarks}

The common message of this article and the preceding Letter \cite{Us} is that the 
``correlation" intrinsic to a fermion state with density operator $\Delta$ may be 
quantified by comparing $\Delta$ to $\quasifree_{\gamma_{\Delta}}$, the 
unique  free state with $1$-particle statistical operator $\gamma_{\Delta}$.  
The correlation we propose to quantify is ``intrinsic" in the sense that it is a 
property solely of the state itself (unlike the ``correlation energy" of a ground state, 
which depends not only upon the state itself, but also upon the Hamiltonian for which it is supposed to be the ground state).

In this article, we have used relative entropy to compare $\Delta$ to $\quasifree_{\gamma_{\Delta}}$, 
but in \cite{Us} we used ``fidelity" instead.  
There we defined  
\begin{equation}
\label{CorrFidelity}
 \mathrm{Corr}(\Delta) = 
- 2\ \log \big( \tr (\Delta^{1/2} \quasifree_{\gamma_{\Delta}} \Delta^{1/2} )^{1/2} \big) \ .
\end{equation}
That is a very reasonable choice, because ``fidelity" has some nice technical properties, 
thanks to which (i) $\mathrm{Corr}$ is monotone and additive just as $\Cor$ is, (ii) $\mathrm{Corr}(\Delta)$ is always finite, and 
(iii) $\mathrm{Corr}$ is a continuous functional when restricted to the domain of $n$-electron states.
In contrast, $\Cor$ is not continuous, and sometimes equals $+\infty$.  
Nonetheless, we prefer $\Cor$ to $\mathrm{Corr}$ because $\Cor$ enjoys 
the superadditivity property expressed in Proposition~\ref{superadditivity}, 
and because the nonfreeness of a pure state equals its (particle-hole symmetric) correlation entropy.

Although our discussion here and in \cite{Us} is limited to density operators 
in the class $\DD$, we remark that nonfreeness and similar measures of fermion 
correlation can be extended to {\it infinite} systems as well.  
It is just a question of comparing two states.  
States of an infinite system are represented by positive linear functionals 
on the CAR algebra over the $1$-particle Hilbert space $\HH$.  
If $\omega$ is a state of the CAR algebra over $\HH$, then there exists a 
unique generalized free state $\gamma_{\omega}$ such that 
$\omega(a^{\dagger}_g a_h) = \gamma_{\omega}(a^{\dagger}_g a_h)$ 
for all $g,h \in \HH$, and $\Cor(\omega)$ may be defined as the entropy 
of $\omega$ relative to $\gamma_{\omega}$ in the sense of 
Umegaki and Araki \cite{OhyaPetz}.  
Similarly, $\mathrm{Corr}(\omega)$ may be defined as the negative logarithm 
of Uhlmann's  transition probability \cite{Bures,Uhlmann} 
connecting these states, which equals (\ref{CorrFidelity}) 
if $\omega(X)=\tr(\Delta X)$.    Bogoliubov automorphisms of the CAR algebra 
will leave $\Cor$ and $\mathrm{Corr}$ invariant.  
Particle-hole duality is implemented by a Bogoliubov transformation, 
and this explains why formula (\ref{CorPure}) is symmetric in $p_j$ and $1-p_j$.  

According to this point of  view, a BCS ground state, which models the condensed state in the theory of 
conventional superconductivity, must be an uncorrelated state, because it is a generalized free state.    
If ``electron correlation" means ``nonfreeness," then conventional 
superconductivity is {\it not} an example of a correlation phenomenon.  

\bigskip

\noindent {\bf Acknowledgements.} 
{\it This work was supported by
the Austrian Ministry of Science (bm:bwk) via its grant for the
Wolfgang Pauli Institute and by the Austrian Science Foundation
(FWF) via the START Project (Y-137-TEC) and by the City of Vienna
Science and Technologie Fund (WWTF) via the project 
``Mathematik und ..." MA-45.}
%

\newpage

\noindent {\bf Appendix A: Proof of Proposition  \ref{p1}  }

Suppose that $\Delta \in \DD$ and (\ref{piaoliang}) holds.  
We want to show that 
\[
S( \Delta | \quasifree_{\gamma_{\Delta}})  \equals - \sum p_j \log p_j - \sum (1-p_j) \log(1-p_j)  - S(\Delta)\ .
\]
To do this we will use the following formula for the relative entropy of two density operators $D$ and $G$, 
which is valid if $\ker(G) \subset \ker(D)$ or, equivalently, if $\overline{\mathrm{Range}(D)} \subset \overline{\mathrm{Range}(G)}$.  
In such cases the entropy of $D$ relative to $G$ is given by
\begin{equation}
\label{lindblad}
      S( D | G ) \equals \sum_{j,k} |\langle \phi_j,\psi_k \rangle|^2 ( d_j \log d_j - d_j \log g_k + g_k - d_j) \ , 
\end{equation}
where $\{\phi_j\}$ and $\{\psi_k\}$ are two orthonormal bases of $\overline{\mathrm{Range}(G)}$ consisting of eigenvectors of $D$ and $G$, respectively, 
and $d_j$ and $g_k$ denote the respective eigenvalues \cite{Lindblad73}.   
The terms in the sum on the right hand side of (\ref{lindblad}) are all nonnegative, so $S( D | G )$ is a finite nonnegative number or $+\infty$.  
Formula (\ref{lindblad}) implies that 
\begin{equation}
\label{blad}
     0 \ \le \ S(D) + S( D | G ) \equals  - \tr( D \log G )
\end{equation}
if $ - \tr( D \log G ) < \infty$. 
In the next paragraph we show that $ \ker(\quasifree_{\gamma_{\Delta}})  \subset \ker(\Delta) $, so we may use formula 
(\ref{blad}) with  $\Delta$ for $D$ and $ \quasifree_{\gamma_{\Delta}} $ for $G$.  It follows that  
\[  S(\Delta|\quasifree_{\gamma_{\Delta}}) = -\tr( \Delta \log \quasifree_{\gamma_{\Delta}} ) - S( \Delta)
\]
 if $-\tr( \Delta \log \quasifree_{\gamma_{\Delta}} ) < \infty$.   In the last paragraph of this proof, we show that 
$-\tr( \Delta \log \quasifree_{\gamma_{\Delta}} ) = - \sum p_j \log p_j - \sum (1-p_j) \log(1-p_j)  $.  That will prove statement (\ref{meili}).

First we establish that $ \ker(\quasifree_{\gamma_{\Delta}})  \subset \ker(\Delta) $.  Let 
$\{g_j\}_{j\in \NNN}$ be an orthonormal basis of $\HH$ such that each $g_j$ lies either in $\ker(\gamma_{\Delta})$ or in $\overline{\mathrm{Range}(\gamma_{\Delta})}$, and let 
\begin{eqnarray}
      K & = & \{ j \in \NNN: g_j \in \ker(\gamma_{\Delta}) \}     \nonumber \\
      R & = & \{ j \in \NNN: g_j \in \overline{\mathrm{Range}(\gamma_{\Delta})} \} \ .  \label {KandR}
\end{eqnarray}
Set $a_j \equiv  a_{g_j}$, and define the operators $a^{\dagger}_{\bs}$ as above for finite subsets $\bs =\{s_1,\ldots,s_n\} \subset \NNN$.  
It is clear from the construction of $\quasifree_{\gamma_{\Delta}}$ by (\ref{spectralQuasifree}) and (\ref{weightQuasifree}) that 
\begin{equation}
   \ker (\quasifree_{\gamma_{\Delta}}) \equals \overline{\mathrm{span}}\{a^{\dagger}_{\bs}\Omega : \bs \cap K \ne \oslash \} \ .
   \label{supportQuasifree} 
\end{equation}
But the following argument shows that each $a^{\dagger}_{\bs}\Omega$ such that $\bs \cap K \ne \oslash$ is also in the kernel of $\Delta$, whence we may conclude $ \ker(\quasifree_{\gamma_{\Delta}})  \subset \ker(\Delta) $.  Given $\bs$ such that $\bs \cap K \ne \oslash$, take $k \in \bs \cap K$, and note that $a^{\dagger}_{\bs}a_{\bs} < a^{\dagger}_k a_k$ holds for the Hermitian projectors $a^{\dagger}_{\bs}a_{\bs}$ and $a^{\dagger}_k a_k$ because $k \in \bs$.  It follows that
\[
    0 \ \le \ \langle  a^{\dagger}_{\bs} \Omega | \Delta  a^{\dagger}_{\bs} \Omega   \rangle \  \le \  \tr( a_{\bs} \Delta a^{\dagger}_{\bs} ) 
      \equals \tr(  \Delta a^{\dagger}_{\bs} a_{\bs} )  \\
     \  \le  \  \tr(  \Delta a^{\dagger}_k a_k )    
     \equals  \langle g_k |   \gamma_{\Delta} g_k \rangle \equals 0\ .
\]
This proves that $a^{\dagger}_{\bs} \Omega \in \ker(\Delta)$ since $\Delta$ is a positive semidefinite operator.

Finally, we verify that $-\tr( \Delta \log \quasifree_{\gamma_{\Delta}} ) = - \sum p_j \log p_j - \sum (1-p_j) \log(1-p_j)  $.  The operator 
$ \log \quasifree_{\gamma_{\Delta}} $ restricted to $\ker( \quasifree_{\gamma_{\Delta}} )$ is supposed to be the zero operator.  Accordingly, we may calculate 
$   \tr( \Delta \log \quasifree_{\gamma_{\Delta}} ) $
with respect to any orthonormal basis of $\overline{\mathrm{Range}(\quasifree_{\gamma_{\Delta}})}$.
We use the basis $\{ a^{\dagger}_{\bs}\Omega : \bs \subset R \}$, where $R$ is as defined in (\ref{KandR}), the basis $\{g_j\}$ of $\HH$ having been chosen so that $\gamma_{\Delta} (g_j) = p_j g_j$.  For a finite subset $\bs \subset \NNN$, define $p(\bs)$ by formula (\ref{weightQuasifree}) and express  
\[
    \quasifree_{\gamma_{\Delta}} = \sum_{\bs \subset R}    p(\bs) \ |  a^{\dagger}_{\bs}\Omega \rangle \langle a^{\dagger}_{\bs}\Omega |
\]
as a sum over finite subsets of $R$ (cf. formula (\ref{spectralQuasifree})).  With this notation, we calculate 
\begin{eqnarray}
\label{calc1}
   \tr( \Delta \log \quasifree_{\gamma_{\Delta}} )  & = & 
    \sum_{\bs \subset R}   \log p(\bs) \langle  a^{\dagger}_{\bs}\Omega | \Delta a^{\dagger}_{\bs}\Omega \rangle \nonumber \\
    & = & 
\sum_{j \in R} \log p_j \sum_{\bs \subset R: j \in \bs} \langle  a^{\dagger}_{\bs}\Omega | \Delta a^{\dagger}_{\bs}\Omega \rangle 
\plus
\sum_{j \in R} \log (1-p_j)  \sum_{\bs \subset R: j \notin \bs} \langle  a^{\dagger}_{\bs}\Omega | \Delta a^{\dagger}_{\bs}\Omega \rangle \ .
\end{eqnarray} 
By (\ref{1PDM}), $   \tr(\Delta a^{\dagger}_j a_j)  =  \langle g_j | \gamma_{\Delta} g_j \rangle = p_j$, and therefore, if $j \in R$, then 
\begin{eqnarray}
\label{calc2}
   \sum_{\bs \subset R: j \in \bs} \langle  a^{\dagger}_{\bs}\Omega | \Delta a^{\dagger}_{\bs}\Omega \rangle  
    & = &
    \sum_{\bs \subset R: j \notin \bs} \langle  a^{\dagger}_{\bs}\Omega | a_j \Delta a^{\dagger}_j a^{\dagger}_{\bs}\Omega \rangle  
    \equals 
   \sum_{\bs \subset R} \langle  a^{\dagger}_{\bs}\Omega | a_j \Delta a^{\dagger}_j a^{\dagger}_{\bs}\Omega \rangle 
   \nonumber \\
    & = &
    \tr(a_j \Delta a^{\dagger}_j) 
   \equals 
   \tr(\Delta a^{\dagger}_j a_j)  
   \equals
   p_j\ .
\end{eqnarray}
Equations (\ref{calc1}) and (\ref{calc2})  imply that 
$
   \tr( \Delta \log \quasifree_{\gamma_{\Delta}} )  =
\sum\limits_{j\in R} p_j \log p_j 
\plus
\sum\limits_{j\in R} (1-p_j) \log (1-p_j)  $.

\bigskip

\noindent {\bf Appendix B: Proof of Corollary \ref{c1}  }

Let $g$ and $h$ be any vectors in $\HH_1$ and $\HH_2$, respectively.  
With respect to the isomorphism (\ref{isomorphism}), the operator $a^{\dagger}_g$ is identified with 
$a^{\dagger}_g \otimes I_2$, where $I_2$ denotes the identity operator on $\FF_2$ and $a^{\dagger}_g$ denotes the creator of $g$ on the Fock space $\FF_1$ (with a slight abuse of notation).  Similarly, $a_h$ is identified with $I_1\otimes a_h$.  Thus, the operator $\Delta a^{\dagger}_g a_h$ is identified with the operator 
$
    \Delta (a^{\dagger}_g \otimes  a_h)
$
on $\FF_1 \otimes \FF_2$.  Let $P:\FF_1 \longrightarrow \FF_1$ denote the projector onto the 
$n$-particle subspace of $\FF_1$.   The hypothesis of the corollary means that $\Delta = (P\otimes I_2)\Delta (P\otimes I_2)$, where 
$P$ denotes the projector onto the $n$-electron space in $\FF_1$.  Thus, $ \tr(\Delta a^{\dagger}_g  a_h)$ equals 
\[
   \tr\big(
   (P\otimes I_2)   \Delta (P\otimes I_2)(a^{\dagger}_g  \otimes a_h)
   \big) 
   \equals \tr\big(
      \Delta (P\otimes I_2)(a^{\dagger}_g  \otimes a_h) (P\otimes I_2)
   \big) 
   \equals \tr\big(\Delta (P a^{\dagger}_g P  \otimes a_h)\big) \ .
\]
But this implies that $\tr(\Delta a^{\dagger}_g  a_h)  = 0$, since $P a^{\dagger}_g P$ is the zero operator on $\FF_1$.  
From the defintion (\ref{1PDM}) of $\gamma_{\Delta}$, we see that $\langle h | \gamma_{\Delta} g \rangle = 0$ when $g \in \HH_1$ and $h \in \HH_2$.  This proves that $\gamma_{\Delta}$ has a direct sum decomposition with respect to the subspaces $ \HH_1$ and $ \HH_2$, and the corollary now follows from Proposition~\ref{superadditivity}.  

\bigskip

\noindent {\bf Appendix C: Nonfreeness formula for $2$-dimensional $1$-particle spaces }

Suppose $\HH$ is a two-dimensional Hilbert space and $\Delta$ is a density operator on $\FF(\HH)$ that commutes with the number operator.  Let $p_1$ and $p_2$ be the natural occupation probabilities and let $q$ be the probability that there are two particles.  
Then 
\begin{eqnarray}
\label{rank 2}
      \Cor(\Delta) & = &   - p_1 \log p_1  -  (1-p_1) \log(1-p_1)  - p_2 \log p_2 - (1-p_2) \log(1-p_2)
     \nonumber \\
      &   &      
      +\ q\log q + (p_1-q)\log(p_1-q) + (p_2-q)\log(p_2-q) \nonumber \\
      &   & +\ (1-p_1-p_2 + q)\log(1-p_1-p_2 + q)\  .
\end{eqnarray}


\begin{thebibliography}{X}



\bibitem{Kutzelnigg}  W. Kutzelnigg, G. Del Re and G. Berthier.  
{\em Correlation coefficients for electronic wave functions}, 
Phys. Rev. 172 (1968) 49 - 59.


\bibitem{LichtnerGriffin}   P. C. Lichtner and J. J. Griffin.  
{\it Evolution of a quantum system: lifetime of a determinant}, 
Phys. Rev. Lett. 37 (23) (1976) 1521 - 1524.



\bibitem{Ziesche}
P. Ziesche. {\em Correlation strength and information entropy},
International Journal of Quantum Chemistry 56 (1995) 363 - 369.


\bibitem{GrobeRzazewskiEberly}
R. Grobe, K. Rzazewski and J.H. Eberly. {\em Measure of
electron-electron correlation in atomic physics}, J. Phys. B, 27
(1994) L503 - L508.



\bibitem{EsquivelEtAl}
R. O. Esquivel, A. L. Rodriguez, R.P Sagar, M. H\^o, and V. H.
Smith. {\em Physical interpretation of information entropy:
numerical evidence of the Collins conjecture}, Phys. Rev.A 54 (1)
(1996) 259 - 265.

\bibitem{GersdorfEtAl}
P. Gersdorf, W. John, J. P. Perdew, and P. Ziesche. {\em
Correlation entropy of the $\mathrm{H}_2$ molecule}, Intl. J. of
Quantum Chemistry 61 (1997) 935 - 941.





\bibitem{ZiescheEtAl}
P. Ziesche, O. Gunnarsson, W. John, and H. Beck. {\em Two-site
Hubbard model, the Bardeen-Cooper-Schrieffer model, and the
concept of correlation entropy}, Phys. Rev. B 55 (16) (1997) 10270
- 10277.



\bibitem{ZiescheEtAl2}
P. Ziesche, V. H. Smith Jr., M. H\^o, S. Rudin, P. Gersdorf, and
M. Taut. {\em The $\mathrm{He}$ isoelectronic series and the Hooke's law
model: correlation measures and modifications of the Collins
conjecture}, Journal of Chemical Physics 110 (13) (1999) 6135 -
6142.



\bibitem{GadreEtAl} S. R. Gadre, S. B. Sears, S. J. Chakravorty, and R. D. Bendale.  
{\it Some novel characteristics of atomic information entropies}, 
Phys. Rev. A 32 (5) (1985) 2602 - 2606.

\bibitem{GuevaraSagarEsquivel}  N. L. Guevara, R. P. Sagar, and R. O. Esquivel. 
{\it Shannon-information entropy sum as a correlation measure in atomic systems}, 
Phys. Rev. A 67 (2003) 012507.

\bibitem{SagarGuevara} R. P. Sagar and N. L. Guevara. 
{\it Mutual information and electron correlation in momentum space},  
Journal of Chemical Physics 124 (2006) 143101.

\bibitem{Gori-GiorgiZiesche} 
P. Gori-Giorgi and P. Ziesche. {\it Momentum distribution of the uniform electron gas: improved parametrization and exact limits of the cumulant expansion}, Phys. Rev. B, 66 (2002) 235116.  [Especially, formula (22) of this article.] 

\bibitem{Us} A. D. Gottlieb and N. J. Mauser. 
{\em New measure of electron correlation}, 
Phys. Rev. Lett., 95 (12): 123003 (2005).


\bibitem{AlickiFannes} R. Alicki and M. Fannes.  
{\it Quantum Dynamical Systems}.
Oxford Univ. Press, Oxford, 2001.


\bibitem{OhyaPetz}   M. Ohya and D. Petz.  
{\it Quantum Entropy and Its Use}. Springer-Verlag, Berlin, 1993.




\bibitem{Bures}
D. Bures. {\em An extension of Kakutani's theorem on infinite
product measures to the tensor product of semifinite W*-algebras},
Trans. Am. Math. Soc. 135 (1969) p.~199.

\bibitem{Uhlmann}
A. Uhlmann. {\em The ``transition probability" in the state space
of a *-algebra}, Reports on Mathematical Physics 9 (1976) 273 -
279.





\bibitem{Lindblad73}  G. Lindblad.  
{\it Entropy, information and quantum measurements}.  
Comm. Math. Phys. 33 (1973) 305 - 322.

\bibitem{Lindblad74}  G. Lindblad.  
{\it Expectations and entropy inequalities for finite quantum systems}.  
Comm. Math. Phys. 39 (1974) 111 - 119.


\bibitem{Lindblad75}  G. Lindblad.  
{\it Completely positive maps and entropy inequalities}.  
Comm. Math. Phys. 40 (1975) 147 - 151.


\bibitem{Uhlmann77}
A. Uhlmann. 
{\em Relative entropy and the Wigner-Yanase-Dyson-Lieb concavity in an interpolation theory},
 Comm. Math. Phys. 54 (1977) 21 - 32.


\bibitem{Petz}  D. Petz {\it Monotonicity of quantum relative entropy revisited}.  
Rev. Math. Physics 15 (2003) 79 - 91.


\bibitem{Fazekas}  P. Fazekas.   
Lecture Notes on Electron Correlation and Magnetism.   
World Scientific, Singapore, 1999.

\bibitem{Anderson} P. W. Anderson.  
{\it New approach to the theory of superexchange interactions}.  
Phys. Rev. 115 (1959) 2 - 13.

\bibitem{KleinSeitz}  D. J. Klein and W. A. Seitz.   
{\it Perturbation expansion of the linear Hubbard model}.   
Phys. Rev. B 8 (1973) 2236 - 2247.




\bibitem{Kato}  T. Kato.  Perturbation Theory for Linear Operators.  
Springer Verlag, Berlin, 1995.  (See Chapter II.)

\end{thebibliography}
\end{document}